\documentclass[10pt,a4paper]{article}
\RequirePackage{amsmath,amssymb}
\RequirePackage[dvipsnames,usenames]{color}
\usepackage{cite}
\usepackage{fullpage}
\usepackage[british]{babel}
\usepackage[latin1]{inputenc}
\usepackage[T1]{fontenc}
\usepackage[final]{showkeys} 
\usepackage[bookmarks]{hyperref}
\usepackage{amsthm}
\usepackage{graphicx}
\usepackage{subfigure}
\usepackage{braket}
\usepackage{mathrsfs}
\usepackage{bm}
\usepackage{amsfonts}
\usepackage{braket}
\newcommand{\be}{\begin{eqnarray}}
\newcommand{\ee}{\end{eqnarray}}
\renewcommand{\d}{\mbox{${\rm d}$}} 
\newcommand{\lp}{\ell_{\rm p}}
\newcommand{\mpl}{m_{\rm p}}
\newcommand{\gn}{G_{\rm N}}

\newcommand{\Rh}{R_{\rm H}}

\title{Is de~Sitter space always excluded in semiclassical $f(R)$ gravity?}
\author{
Roberto~Casadio$^{ab}$\thanks{E-mail: casadio@bo.infn.it},
Andrea~Giugno$^{c}$\thanks{E-mail: A.Giugno@physik.uni-muenchen.de},
Andrea~Giusti$^{d}$\thanks{E-mail: agiusti@ubishops.ca}
$\,$
and
Valerio~Faraoni$^d$\thanks{E-mail: vfaraoni@ubishops.ca}
\\
\\
$^a${\em Dipartimento di Fisica e Astronomia, Universit\`a di Bologna}
\\
{\em via Irnerio~46, I-40126 Bologna, Italy}
\\
\\
$^b${\em I.N.F.N., Sezione di Bologna, I.S.~FLAG}
\\
{\em via B.~Pichat~6/2, I-40127 Bologna, Italy}
\\
\\
$^c${\em Arnold Sommerfeld Center, Ludwig-Maximilians-Universit\"at}
\\
{\em Theresienstra{\ss}e 37, 80333 M\"unchen, Germany}
\\
\\
$^d${\em Department of Physics and Astronomy, Bishop's University,}
\\
{\em 2600 College Street, Sherbrooke Qu\'ebec, Canada J1M 1Z7}
}
\begin{document}
\maketitle
\begin{abstract}
It is shown that the recent corpuscular description of gravity generically 
excludes de~Sitter spacetime in any semiclassical version of $f(R)$ 
gravity. A phantom phenomenology of the cosmic dynamics is also 
naturally excluded.
\end{abstract}

\newpage

\section{Introduction}
\label{intro}

The inflationary scenario, whose original formulation was initially 
advocated by Starobinsky~\cite{starobinskyI-1, starobinskyI-2} and Guth~\cite{Guth-first}, 
finds its fundamental motivation in the quest for the resolution of three 
major conceptual issues in early universe cosmology, namely the 
homogeneity and flatness of our universe, and the magnetic-monopole 
problem. Later on, Linde~\cite{Linde}, Albrecht, and Steinhardt~\cite{AS} 
built on this proposal introducing a scheme in which the accelerated 
expansion was driven by a scalar field, commonly referred to as the
{\em inflaton}, rolling slowly on a plateau of the potential toward its  
minimum.
Scalar field inflation has the added value of providing a 
mechanism for the generation of density perturbations that constitute  
the seeds for structures that grow in the later dust-dominated era.
The key idea behind scalar field inflation is that, if this  
plateau is sufficiently flat, the phase of (quasi-)exponential expansion 
lasts long enough to solve the cosmological problems mentioned above,  
providing also a mechanism for the universe to {\em gracefully} leave
this highly accelerated phase. 
This last paradigm can be recast as a $f(R)$ theory of gravity
(see, {\em e.g.},~\cite{starobinsky-1,starobinsky-2,starobinsky-3,starobinsky-4}) and it has become 
part of the standard picture of the early universe. 
The Starobinsky model~\cite{starobinskyI-1, starobinskyI-2} appears to be 
particularly favoured by present observations~\cite{planck,WMAP}.
\par
In inflationary theory, it is {common practice to employ 
the semiclassical picture of gravity, which restricts the picture to
the dynamics of quantum fields on a classical curved spacetime.
Further, only quantum fluctuations of the fields around classical values 
are allowed.
However, this approximation inevitably 
fails to capture relevant quantum properties of gravity in the very early
universe~\cite{dvali}.
A way to preserve a quantum mechanical 
description of the early universe consists of conceiving the classical
spacetime geometry as an emergent, rather than a fundamental,
property of nature.
A scheme that successfully implements this 
picture at both astrophysical and cosmological scales is the so-called 
corpuscular theory of gravity~\cite{Gia1-a,Gia1-b,Gia1-c,Gia1-d}. 
Its fundamental idea is that a classical spacetime can be thought of as a 
{\em self-sustained marginally bound state} of a large number of
{\em soft off-shell gravitons}.
The high multiplicity of this state allows one to recover the classical and
semiclassical features of the emergent spacetime.
The first implementation of the corpuscular model~\cite{Gia1-a,Gia1-b,Gia1-c,Gia1-d} investigated
the physics hidden inside a black hole's event horizon.
Precisely,  within this framework a black hole of 
mass $M$ is understood as a marginally bound state of $N$ off-shell 
gravitons of typical Compton length $\lambda _{\rm G}$ of the order of 
the Schwarzschild radius $\Rh = 2 \, \gn \, M$.~\footnote{We shall use units 
with $c=1$ and the Newton constant $\gn=\lp/\mpl$, where $\lp$ and $\mpl$ 
are the Planck length and mass, respectively, and the reduced Planck 
constant  is $\hbar=\lp\,\mpl$.}
Then, even when the system enters the 
strong coupling regime, the requirement of a marginally bound state allows 
us to frame this theoretical set-up as a Newtonian theory of many 
gravitons that appear to be loosely confined in a 
potential well of size $\lambda _{\rm G} \simeq \Rh$.
As a consequence, 
the effective gravitational coupling among the constituent 
gravitons of this bound state scales like $\alpha \sim 1/N$.
This result is of paramount importance for this whole picture since it allows
one to understand classical black holes as the result of bound states of 
gravitons on the verge of a quantum phase transition. 
Indeed, while the 
effective coupling $\alpha$ is very small, the collective coupling for the 
system $g = N \, \alpha \sim 1$, and this configuration clearly matches 
the traditional picture of an interacting Bose-Einstein condensate at the 
critical point.
This ultimately allows us to provide a corpuscular 
interpretation of Hawking radiation in terms of gravitons
leaking out of the bound state~\cite{Gia1-a,Gia1-b,Gia1-c,Gia1-d}. 
Further, working in this general framework, it is possible to recover the correct
post-Newtonian expansion of the gravitational field generated by a static,
spherically symmetric source~\cite{baryons-1,baryons-2} in a fully quantum framework and the 
Bekenstein-Hawking area law~\cite{bekenstein}, including 
semiclassical logarithmic corrections~\cite{planckian-1,planckian-2} for the Hawking 
radiation.
What is more, when applied to the case of maximally symmetric 
spaces, the corpuscular theory of gravity has  proven effective in reproducing
the main features of  inflation~\cite{dvali,cko} and dark matter
phenomenology~\cite{Cadoni:2017evg-1,Cadoni:2017evg-2}.
In particular, it was shown in Refs.~\cite{inflation, GG} that Starobinsky's
inflation is naturally embedded in this framework.
\par
In this work, we show that a corpuscular description of $f(R)$ 
theories of gravity always excludes the exact de~Sitter spacetime,
thus generalizing the results obtained in Ref.~\cite{swampGia},
while providing some (fully quantum mechanical) 
insight on the physically admissible $f(R)$ models.
\section{de~Sitter universe}
\label{sez:cc}
It is well known that the de~Sitter metric is an exact solution of 
the modified theory of gravity~\cite{starobinsky-2,starobinsky-3, Barrow}
\be
S
=
\frac{1}{16\,\pi\,\gn}
\int \d^4 x\,\sqrt{-g}\,f(R)
\ ,
\label{SfR}
\ee
with~\cite{zerbini,lust}
\be
f(R)=\gamma\,\lp^2\,R^2
\ ,
\label{fR2}
\ee
where $\gamma$ is a dimensionless constant.
We recall that the equation of motion arising from the 
variation of the action~\eqref{SfR} for a
spatially flat Friedmann-Lema\^itre-Robertson-Walker (FLRW) metric,
\be
\d s^2=-\d t^2 +a^2(t)\left(\d r^2+ r^2\,\d\Omega^2\right) \,,
\ee
is~\cite{starobinsky-1,starobinsky-2,starobinsky-3,starobinsky-4,zerbini,Odintsov}
\be
6\,f'(R)\,H^2
=
R\,f'(R)-f(R)-6\,H\,\dot R\,f''(R)
\,,
\label{eomR2}
\ee
where a prime and an overdot denote differentiation with respect to $R$
and to the comoving time $t$, respectively. 
In particular, for the theory~\eqref{fR2}, one obtains
\be
12\,R\,H^2 = R^2 -12\,H\,\dot R  \label{eom2}
\ee
and, for de~Sitter spacetime with $a(t)=\mbox{e}^{\sqrt{\Lambda/3}\,t}$ 
and constant $H\equiv \dot{a}/a=\sqrt{\Lambda /3}$, one  has
\be
\dot R = 24\,H\,\dot H =0  \,.
\ee
and
\be
\frac{R}{12}
=
\frac{\dot{H}+2\, H^2}{2}
=
H^2
=
\frac{\Lambda}{3}
\ . 
\label{fried1}
\ee
Integrating the left hand side of Eq.~\eqref{fried1} over a sphere 
of 
Hubble radius 
$L_\Lambda=H_\Lambda^{-1} = \sqrt{ 3/\Lambda}$ yields
\be
L_\Lambda^3\,H^2_\Lambda
\simeq
L_\Lambda  \label{n}
\equiv
-\gn\,U_{\rm N}
\ee
and, likewise, the right hand side of Eq.~\eqref{fried1} yields
\be
L_\Lambda^3
\left({\Lambda}/{3}\right)
\simeq
L_\Lambda
\equiv
\gn\,U_{\rm PN}
\,,
\label{pn}
\ee
where we introduced a ``Newtonian'' and a ``post-Newtonian'' (or post-Minkowskian)
energy $U_{\rm N}$ and $U_{\rm PN}$, respectively.\footnote{Factors of 
order unity will be often omitted from now on.}
These expressions will be our starting point to investigate the 
relation between corpuscular model and de~Sitter space. 
\section{Corpuscular de~Sitter and depletion}
\label{sez:depletion}
%
Let us begin by assuming that matter  and the corpuscular state of 
gravitons together reproduce the Friedmann equation of 
cosmology,  which we write as the Hamiltonian constraint
\be
\mathcal{H}_{\rm M}+\mathcal{H}_{\rm G} = 0
\ ,
\label{H0}
\ee
where $\mathcal{H}_{\rm M}$ is the matter energy and $\mathcal{H}_{\rm G}$
is the analogue quantity for the graviton state.
Local (Newton or Einstein) gravity being attractive implies that $\mathcal{H}_{\rm G}\le 0$,
although this is not true for the graviton 
self-interaction~\cite{baryons-1,baryons-2}, and might not be true
for the cosmological condensate of gravitons as a whole, as we are now 
going to discuss.
\par
In order to obtain the de~Sitter spacetime in general relativity, that is the theory~\eqref{SfR}
with $f(R)=R$, one must add a cosmological constant 
term, or vacuum
energy density $\rho_\Lambda$, so that the Friedmann equation,
\be
3\,H^2
=
8\,\pi\,\gn\,\rho_\Lambda
\,, \label{frwDS}
\ee
equals precisely Eq.~\eqref{fried1}.
Upon integrating again on the volume inside the Hubble radius, one has
\be
L_\Lambda 
\simeq
\gn\,L_\Lambda^3\,\rho_\Lambda
\simeq
\lp\,\frac{M_\Lambda}{\mpl}
\,,
\label{LgM}
\ee
which looks exactly like the expression of the horizon radius for a black 
hole of mass $M_\Lambda$, and is the reason why it was conjectured that de~Sitter
spacetime could likewise be viewed as a graviton condensate~\cite{dvali}.
\par
One can roughly describe the corpuscular model by assuming 
that the (soft, virtual) graviton self-interaction gives rise to a 
condensate of $N_\Lambda$ gravitons of typical Compton length
$\lambda\simeq L_\Lambda$, so that $ 
M_\Lambda=N_\Lambda\,{\lp\,\mpl}/{L_\Lambda}$,
and the usual consistency condition
\be
M_\Lambda
\sim
\sqrt{N_\Lambda}\,\mpl \label{MN}
\ee
for the graviton condensate immediately follows from Eq.~\eqref{LgM}.
Equivalently, one finds
\be
L_\Lambda
\sim
\sqrt{N_\Lambda}\,\lp
\,,
\label{LMn}
\ee
which shows that for a macroscopic universe one needs $N_\Lambda\gg 1$.  
The above relations do not need to hold for gravitons that do not belong to 
the condensate, therefore one expects deviations to occur if regular 
matter is added~\cite{Cadoni:2017evg-1,Cadoni:2017evg-2}, or if the system is driven out of 
equilibrium.
\par
We can refine the above corpuscular description of de~Sitter
 by following the line of reasoning of Refs.~\cite{baryons-1,baryons-2}, where it was 
shown that the maximal packing condition yielding  
the scaling relation~\eqref{LMn} for a black hole actually follows 
from the 
energy balance~\eqref{H0}
when matter becomes totally negligible.
In the present case, matter is absent {\em a priori\/} and 
$\mathcal{H}_{\rm M}=0$, so that one is left with
\be
\mathcal{H}_{\rm G}
\simeq
U_{\rm N}
+
U_{\rm PN}
=
0
\,.
\label{H00}
\ee
The negative ``Newtonian energy'' of the $N_\Lambda$ gravitons can be obtained 
from a coherent state description of the condensate~\cite{baryons-1,baryons-2} in which each graviton has negative
binding energy $\varepsilon_\Lambda$ given by the Compton relation, that is
\be
U_{\rm N}
\simeq
M_\Lambda\,\phi_{\rm N}
=
N_\Lambda\,\varepsilon_{\Lambda}
=
-N_\Lambda\,\frac{\lp\,\mpl}{L_\Lambda}
\,.
\label{UN0}
\ee
The positive ``post-Newtonian'' contribution is then given by the (bootstrapped) graviton
self-interaction term~\cite{baryons-1,baryons-2}
\be
U_{\rm PN}
\simeq
N_\Lambda\,\varepsilon_\Lambda\,\phi_{\rm N}
=
N_\Lambda^{3/2}\,\frac{\lp^2\,\mpl}{L_\Lambda^2}
\,,
\label{UPN0}
\ee
where we used the Newtonian potential 
\be
\phi_{\rm N}
=
-\frac{N_\Lambda\,\lp\,\mpl}{M_\Lambda\,L_\Lambda}
=
-\sqrt{N_\Lambda}\,\frac{\lp}{L_\Lambda}
\,,
\ee
as follows from Eq.~\eqref{UN0} and the scaling relation~\eqref{MN}.

In an ideal de~Sitter universe, gravitons should satisfy the balance condition~\eqref{H00}.
Let us rewrite the Hamiltonian~\eqref{H00} as
\be
\mathcal{H}_{\rm G}^{(2)}
\simeq
\beta
\left(
U_{\rm N}
+
U_{\rm PN}
\right)
\,,
\label{H2}
\ee
corresponding to the effective metric action~\eqref{SfR} with Eq.~\eqref{fR2}. 
Here we have introduced the dimensionless parameter $\beta>0$ of order unity 
in order to keep track of this contribution.
Let us include a term corresponding to the Einstein-Hilbert action, 
that is
\be
\mathcal{H}_{\rm G}^{(1)}
\simeq
\alpha\,
U_{\rm N}
\,,
\label{H1}
\ee
where $\alpha>0$.
The full energy balance is therefore
\be
\mathcal{H}_{\rm G}
= 
\mathcal{H}_{\rm G}^{(1)}
+
\mathcal{H}_{\rm G}^{(2)}
\simeq
\left(\alpha+\beta\right)
U_{\rm N}
+
\beta\,U_{\rm PN}
= 0 
\label{H12}
\ee 
and, because of the term proportional to $\alpha$, we expect the expressions~\eqref{n}
and~\eqref{pn} for the ideal de~Sitter condensate to no longer constitute a solution.
In fact, we are interested in a stage when departures from the de~Sitter scalings are small, and
we can therefore assume that the potentials now take the slightly more 
general form
\be
\gn\,U_{\rm N}
\simeq
-L^3\,H^2
\label{n2}
\ee
and
\be
\gn\,U_{\rm PN}
\simeq
L^3\,L_\Lambda^{-2}
\,, 
\label{pn2}
\ee
where $L\sim L_\Lambda$ is the new Hubble radius.
Substitution into Eq.~\eqref{H12} yields 
\be
L^3
\left[
-\left(\alpha+\beta\right)
H^2
+
\beta\,L_\Lambda^{-2}
\right]
\simeq
0
\,,
\label{H12app}
\ee
which is solved by
\be
H^2
\simeq
\frac{\beta}{\alpha+\beta}\,
\frac{1}{L_\Lambda^2} \,.
\ee
Of course, the de~Sitter case is properly recovered when $\alpha=0$, but
$\alpha>0$ implies that $H<H_\Lambda$ as expected. 
If the system starts with $H \simeq H_\Lambda$, the 
time derivative $\dot H$ must be negative
({\em i.e.}, the universe does not superaccelerate) 
in order to ensure the constraint~\eqref{H12} holds at all times.
This can be seen explicitly by writing
\be
H = H_\Lambda + \dot H\,\delta t
\,,
\ee
where the typical time scale $\delta t\simeq L_\Lambda$, since gravitons 
of Compton length
$L_\Lambda$ cannot be sensitive to shorter times.
Equation~\eqref{H12app} finally yields
\be
\dot H
\simeq
-\frac{\alpha}{\alpha+\beta}\,
\frac{H_\Lambda}{\delta t}
\simeq
-\frac{\alpha}{\alpha+\beta}\,
\frac{1}{L_\Lambda^2}
\,.
\label{dotHc}
\ee
Further, the slow-roll parameter in the corpuscular model is 
\be
\epsilon
\equiv
-\frac{\dot H}{H^2}
\simeq
\frac{\alpha}{\alpha+\beta}
\ ,
\label{eps} 
\ee
and one obtains $ \epsilon = 0 $ in the limit $\alpha\to 0$,
provided the quantum depletion can be neglected.
\par
In detail, gravitons in the condensate generate the effective
Hubble parameter $H\sim N_\Lambda^{-1/2}\sim L_\Lambda^{-1}$,
but they also scatter and deplete.
Their number therefore changes in time, according to \cite{dvali,cko} 
\be
\frac{\lp\,\dot N_\Lambda}{\sqrt{N_\Lambda}}
=
\left.
\frac{\lp\,\dot N_\Lambda}{\sqrt{N_\Lambda}}
\right|_{\rm eom}
+
\left.
\frac{\lp\,\dot N_\Lambda}{\sqrt{N_\Lambda}}
\right|_{\rm q}
\ ,
\ee
where the classical equation of motion gives (for $\alpha\not=0$) 
\be
\left.
\frac{\lp\,\dot N_\Lambda}{\sqrt{N_\Lambda}}
\right|_{\rm eom}
\simeq
-\frac{\dot H}{H^2}
\simeq
\frac{\alpha}{\alpha+\beta}
\,,
\ee
and the purely quantum depletion yields
\be
-\frac{\dot H}{H^2}
=
\lp\,\dot M_\Lambda
\simeq
-
\frac{\mpl^2}{M_\Lambda^2}
\simeq
-\lp^2\,H^2
\simeq
-\frac{\beta}{\alpha+\beta}\,\frac{1}{N_\Lambda}
\ .
\quad
\ee
Putting the two terms together we obtain, for $\alpha\ll\beta\simeq 1$, 
\be
-\frac{\dot H}{H^2}
\simeq
\frac{\lp\,\dot N_\Lambda}{\sqrt{N_\Lambda}}
\simeq
\alpha
\left(
1
-
\frac{\beta}{\alpha\,N_\Lambda}
\right)
\ ,
\label{dotN}
\ee
leading to a critical value of $\alpha$
\be
\alpha
\gtrsim
\alpha_{\rm c}
\simeq
\frac{\beta}{N_\Lambda}
\sim
\frac{\lp^2}{L_\Lambda^2}
\ ,
\label{Ac}
\ee
which can be interpreted as a minimum ``distance'' from de~Sitter space.
\par
Now, the Hamiltonian ${\cal H}_{G}$ for a polynomial $f(R)$ theory of gravity 
always consists of two contributions ${\cal H}_{G}^{(1)}$ and ${\cal H}_{G}^{(2)}$
in corpuscular gravity.
Effectively, the universe moves away from a  de Sitter fixed point of Eq.~\eqref{fR2},
$f(R)\simeq R^2$, and approaches asymptotically another fixed point of general relativity with positive
cosmological constant, that is $f(R)\simeq R-\Lambda$~\cite{inflation, 
Cadoni:2017evg-1,Cadoni:2017evg-2}.   
More precisely, assuming that $R\simeq \Lambda$ is approximately 
constant, Eq.~\eqref{eomR2} yields
\be 
6 \, f'(R) \, H^2 
\simeq
R \, f'(R) - f(R)
\,. \label{eomN}
\ee
If we write
\be 
f(R)
=
\sum _{k=1} ^N a_k \,\lp^{2k-2}\, R^k
\,, 
\label{fRsum}
\ee
we find
\be 
R \, f'(R) - f(R)
=
\sum _{k=2} ^N (k-1) \, a_k \, \lp^{2k-2}\, R^k 
\ee
and, up to coefficients of order unity, Eq.~\eqref{eomN} reduces to
\begin{equation}
H^2 
\left( a_1 + \sum _{k=2} ^N k \, a_k \, \lp^{2k-2}\, R^{k-1}\right) 
\simeq
\sum _{k=2} ^N (k-1) \, a_k \, \lp^{2k-2}\, R^k
\,,
\end{equation}
where we singled out the coefficient $a_1$ for later convenience.
Using $R\simeq \Lambda$ and defining $n = k-1$ yield
\be
a_1 \, H^2
&\simeq&
\sum _{n=1} ^{N-1}
\left[n \left( \Lambda - H^2\right)
-H^2
\right]
a_{n+1} \, \lp^{2n}\,\Lambda^{n}
\nonumber 
\\
&\simeq&
\sum _{n=1} ^{N-1}
\left( \Lambda - H^2  \right)
n \, a_{n+1} \, \lp^{2n}\,\Lambda^{n}
\,.
\ee
If we then use Eqs.~\eqref{n} and~\eqref{pn}, it is easy to see that
\be 
H^2 
&\simeq& 
- \frac{\gn \, U_{\rm N}}{L_\Lambda ^3} 
\simeq 
- \frac{\lp}{\mpl \, L_\Lambda ^3} \, U_{\rm N} 
\ee
and
\be
\Lambda 
&\simeq& 
\frac{\gn \, U_{\rm PN}}{L_\Lambda ^3}
\simeq
\frac{\lp}{\mpl \, L_\Lambda ^3} \, U_{\rm PN}
\,,
\ee
which lead to
\be 
a_1 \, U_{\rm N} 
+ 
\left(U_{\rm N} + U_{\rm PN}\right)
\sum _{n=1} ^{N-1} 
n \, a_{n+1} \, \left( \frac{\lp^3 \, U_{\rm PN}}{L_{\Lambda}^3\,\mpl} 
\right)^{n} \simeq 0
\,.
\qquad
\ee
For the Starobinsky model ($N=2$), the latter reads simply
\be 
a_1 \, U_{\rm N} 
+ 
a_{2}
\left( \frac{\lp^3 \, U_{\rm PN}}{L_\Lambda ^3\,\mpl} \right) 
\left(U_{\rm N} + U_{\rm PN}\right) 
\simeq 0
\,.
\ee
Now, recalling that $a_1 = \alpha$ and $a_2 = \gamma$, it is
\be 
\alpha \,U_{\rm N} 
+ 
\gamma\,\frac{\lp^2}{L_\Lambda^2}
\left(U_{\rm N} + U_{\rm PN}\right) 
\simeq
0
\,,
\ee
in which we further used the de~Sitter value~\eqref{pn} for $U_{\rm PN}$,
since we are interested in studying the stability of this space.
Comparing with Eq.~\eqref{H12}, one recovers 
\be
\beta
=
\beta_2
\equiv
\gamma\,\frac{\lp^2}{L_\Lambda^2}
\,.
\ee
Likewise, for the general Eq.~\eqref{fRsum}, one obtains Eq.~\eqref{H12}
with $\alpha=a_1$ and
\be
\beta
=
\beta_N
\equiv
\sum_{n=1}^{N-1} n\,a_{n+1}
\left(\frac{\lp}{L_\Lambda}\right)^{2n}
\,.
\ee
All of the above estimates can be improved but, comparing 
with the previous discussion, we expect again a bound 
similar to the one in 
Eq.~\eqref{Ac}.
%
%
\section{Conclusions}
When the corpuscular model is studied in conjunction with 
quadratic corrections to the Einstein-Hilbert action, as is natural, and 
is applied to early universe cosmology (as in Starobinsky inflation, 
currently favored by cosmological observations), one finds that an exact 
de ~Sitter space is always excluded.
This result is significant because de~Sitter spacetime
is the most common attractor in the FLRW cosmology 
based on general relativity and on $f(R)$ theories of gravity and,
in order to be a fixed point, de~Sitter space must be an {\em exact}
solution of the cosmic dynamics. 
\par
From an effective point of view, in the corpuscular  model a universe 
that is close to (but not exactly) de~Sitter in the high curvature regime where 
$f(R) \propto R^2$ moves away from it due to gravitons leaking out of 
the bound state and approaches a different de~Sitter space in the low curvature
regime where  $f(R) \simeq R-\Lambda$.
If the universe was described at all times by a purely classical $f(R)$ theory,
this beheavior  would be described by a heteroclinic trajectory linking two de~Sitter
fixed points in phase space, but this is not the case in the corpuscular model
with graviton depletion, in which the universe departs from the initial, high curvature,
coherent state. 
\par
From a classical perspective, an $f(R)$ theory can be 
conformally mapped into Einstein gravity minimally coupled to a scalar 
field with a complicated potential, the latter is known as the Einstein 
conformal frame whereas the former defines the Jordan frame for the model. 
A key result of Ref.~\cite{dvali} is that the exclusion of the 
de~Sitter space is a direct consequence of the quantum depletion in the 
Einstein frame. Here, instead, we offer a different angle to this result 
based on  the Hamiltonian constraint of cosmology and the Jordan frame 
description of quantum depletion.
\par
Finally, the cosmological phenomenology of the corpuscolar 
model is quite reasonable and not exotic, in the sense that phantom 
behavior is always avoided since  $\dot H$ must be negative and, in addition
to never reaching a de~Sitter state, the universe never superaccelerates.
Further details of this intriguing corpuscular cosmology 
of the early universe will be reported elsewhere.

\section*{Acknowledgments}

R.C.~is partially supported by the INFN grant FLAG.
This work has also been carried out in the framework of activities of the National Group
of Mathematical Physics (GNFM, INdAM) of Italy.
V.F.~and A.G.~are supported by the Natural Sciences and Engineering Research Council
of Canada (Grant No.~2016-03803 to V.F.) and by Bishop's University.

 \end{document}